\documentclass[aps,pra,twocolumn,superscriptaddress]{revtex4-1}
\usepackage{amsmath}
\usepackage{bm}

\newcommand{\ket}[1]{|#1\rangle}
\newcommand{\meanv}[1]{\left\langle #1 \right\rangle}
\newcommand{\bb}[1]{\left( #1 \right)}

\usepackage[dvipdfmx]{graphicx}
\usepackage{type1cm}
\usepackage[usenames,svgnames]{xcolor}
\usepackage{color}
\usepackage{url}
\usepackage{natbib}
\usepackage[dvipdfmx]{hyperref}
\hypersetup{
     colorlinks=true,       		
     linkcolor=magenta,          	
     citecolor=blue,            
     filecolor=blue,      		
     urlcolor=blue,           	
    runcolor=cyan,
 }

\begin{document}


\title{Superadiabatic generation of cat states in bosonic Josephson junctions \\ under particle losses}


\author{Takuya Hatomura}
\email[The following e-mail address is at least valid until the end of March 2019: ]{hatomura@spin.phys.s.u-tokyo.ac.jp}
\affiliation{Department of Physics, University of Tokyo, 113-8654 Tokyo, Japan}

\author{Krzysztof Paw{\l}owski}
\affiliation{Center for Theoretical Physics, Polish Academy of Sciences, 02-668 Warsaw, Poland}


\date{\today}

\begin{abstract}
We investigate a superadiabatic scheme to produce a cat state in a bosonic Josephson junction in absence and presence of particle losses. 
The generation scheme is based on shortcuts to adiabaticity and strongly relies on the parity conservation. 
The parity conservation also ensures that the produced state is a superposition of cat states with various sizes, i.e., a ``cats state". 
Parity is also the quantity to be measured in order to utilize the produced state in interferometry. 
The generation scheme still works even if a number of particle losses during generation are substantial. 
\end{abstract}


\maketitle

%

\section{Introduction}
Paradigmatic examples of non-classical states would be macroscopic superpositions as proposed by Erwin Schr\"odinger more than 80 years ago. 
The Schr\"odinger cat state is actually one of the most weird phenomena in quantum mechanics, also known by a broad audience, which moreover could be potentially used in quantum computing \cite{Cochrane1999} or interferometry \cite{Demkowicz-Dobrzanski2015, Pezze2018}. 
There are numerous proposals to produce such states. 
On the other hand, not many successful experiments have been reported in this field -- the results have been limited to small systems consisting of a few ions \cite{Leibfried2005,Monz2011} or several photons \cite{Deleglise2008,Vlastakis2013}.

Many ideas to produce the cat states originate in simple theoretical models, which approximately describe certain real systems. 
Those are sometimes over-simplified, however they provide us with a lot of intuition about dynamics of ultracold atoms. 
An illustrative example is bimodal Bose-Einstein condensation typically understood in the frame of the single-mode approximation, where we assume that all atoms share a common spatial mode and their dynamics is limited in two internal levels~\cite{Cirac1998,Steel1998}. 
This is actually a system, for which many schemes to produce highly entangled states have been proposed. 
Not only that but also many successful experiments have been reported, in which the squeezed~\cite{Esteve2008, Gross2010, Riedel2010} and over-squeezed~\cite{Strobel2014} states were produced.

There is a certain class of systems, whose ground states become the cat states for certain parameters. 
Indeed, bimodal Bose-Einstein condensation can be the cat state as the ground state~\cite{Cirac1998}. 
For such a system, we can think of a generation scheme, where we adiabatically drive this system from a parameter region, for which the ground state can be easily obtained, to a parameter region, for which the ground state is a macroscopic superposition. 
The main difficulty of this scheme is to maintain adiabaticity, i.e., critical slowing down disables us for keeping a system in the ground state. 
One of the ways to overcome this difficulty is to utilize shortcuts to adiabaticity~\cite{Torrontegui2013}. 
In this solution, besides dynamical changes of parameters, we add time-dependent terms, which help in preserving adiabaticity of original problems~\cite{Demirplak2003,Demirplak2005,Berry2009}. 
This technique turned out to be very effective in control of Bose-Einstein condensation. 
For example, transformation of trap potential~\cite{delCampo2011b,delCampo2012a,delCampo2013} and generation of highly entangled states~\cite{Julia-Diaz2012,Yuste2013,Opatrny2016,Hatomura2018a} have been proposed. 
Some experimental realizations of shortcuts to adiabaticity in Bose-Einstein condensation have also been reported~\cite{Schaff2011,Schaff2011a,Bason2012}.

For bimodal Bose-Einstein condensation, such additional terms counteracting diabatic changes were proposed~\cite{Hatomura2018a}. 
The form of the additional Hamiltonian is nothing but the two-axis countertwisting (TACT) Hamiltonian~\cite{Kitagawa1993}. 
There are many proposals to implement the TACT Hamiltonian, for example, counter-propagating flows of ultracold atoms on a ring~\cite{Opatrny2015} and dipolar interactions in spinor Bose-Einstein condensation~\cite{Kajtoch2016}, while it has not been realized yet. 
Recently, physically related schemes have been also studied~\cite{Feldmann2018}.

Although realization of the TACT Hamiltonian is future work, one could question if the superadiabatic scheme based on shortcuts to adiabaticity~\cite{Hatomura2018a} has any chance to work. 
In particular, particle losses, which take place due to collisions among atoms and always happen in ultracold atoms, cause various channels of decoherence and destroy macroscopic superpositions. 
Indeed, in the case of the paradigmatic one-axis twisting Hamiltonian, which can be used to generate a macroscopic superposition, the interplay of non-linear dynamics and particle losses induces phase noise,  theta noise, and Gaussian damping~\cite{Spehner2014, pawlowski2013}, and thus current technology limits the maximal cat states up to dozens of atoms~\cite{Pawowski2017}. 
In fact without special tricks, just a single lost atom during  one-axis twisting dynamics can cause huge phase noise smearing the state over the whole Bloch sphere, and therefore all quantum correlations are destroyed.
One can then question if the non-linear terms added to the Hamiltonian, which are needed for shortcuts to adiabaticity, induce side effects under particle losses or not. In particular, it would be of interest if there is any advantage of adding such extra non-linear terms. 
From this viewpoint, here we discuss effects of losses in the superadiabatic scheme~\cite{Hatomura2018a} and indicate possibility to obtain potentially useful macroscopic superpositions.

This paper is constructed as follows. 
Section~\ref{Sec.method} is the brief summary of our methodology. 
We review basic properties of bosonic Josephson junctions and explain a generation scheme based on adiabatic time evolution in Sec.~\ref{Sec.AdGen}. 
We introduce the formalism of shortcuts to adiabaticity by counter-diabatic driving and apply it to a bosonic Josephson junction in Sec.~\ref{Sec.STAgen}. 
We introduce the Monte Carlo wave function method in order to take particle losses into account in Sec.~\ref{Sec.Ploss}. 
Section~\ref{Sec.results} is devoted to results. 
In Sec.~\ref{Sec.gen}, the superadiabatic scheme is compared with the naive adiabatic scheme. 
In Sec.~\ref{Sec.pari}, we discuss the form of a produced state by introducing the parity operator and discuss how to detect it. 
In Sec.~\ref{Sec.dist}, we show that particle losses do not drastically change distributions of states both in the adiabatic and the superadiabatic schemes. 
In Sec.~\ref{Sec.surv}, we discuss a possibility of detecting cat states and survival of entanglement under particle losses. 
We summarize the present work in Sec.~\ref{Sec.sum}. 

%
%
\section{\label{Sec.method}Methods}
%
%
\subsection{\label{Sec.AdGen}Adiabatic scheme generating a cat state}
Suppose that a Bose-Einstein condensate can be well described by using the single-mode approximation for spatial degrees of freedom and two internal levels. 
These two modes are represented by bosonic operators $a_1$ ($a_1^\dag$) and $a_2$ ($a_2^\dag$), respectively. 
By using the angular momentum expression
\begin{eqnarray}
&J_x=\frac{1}{2}(a_1^\dag a_2+a_2^\dag a_1), \label{Eq.AMEx} \\
&J_y=\frac{i}{2}(a_2^\dag a_1-a_1^\dag a_2), \label{Eq.AMEy} \\
&J_z=\frac{1}{2}(a_1^\dag a_1-a_2^\dag a_2), \label{Eq.AMEz}
\end{eqnarray}
the Hamiltonian of a bosonic Josephson junction is given by~\cite{Cirac1998,Steel1998}
\begin{equation}
\mathcal{H}_\mathrm{BJJ}=\hbar\chi J_z^2+\hbar\Omega J_x,
\label{Eq.BJJ}
\end{equation}
where we assume that both the nonlinear interaction parameter $\chi$ and the coupling parameter $\Omega$ are negative, $\chi<0$ and $\Omega<0$. 
Here, the angular momentum operators satisfy the usual commutation relations, $[J_\alpha,J_\beta]=i\epsilon_{\alpha\beta\gamma}J_\gamma$. 
The competition between the nonlinear interaction and the coupling results in the phase transition at the critical point with the parameter $\Lambda\equiv\chi N/\Omega=1$, where $N$ is number of particles. 
The system is in the disordered phase when $0<\Lambda<1$ and in the ordered phase when $\Lambda>1$~\cite{Cirac1998,Botet1982}. 

The ground state in the ordered phase is a cat state, i.e., a macroscopic superposition of a mode-1 condensate and a mode-2 condensate~\cite{Cirac1998,Botet1982}. 
Especially, the ground state corresponding to $\Lambda^{-1}=0$ can be the NOON state~\cite{Lee2002}. 
However, if the system is cooled down into the ground state directly, it would result in a statistical mixture of a mode-1 condensate and a mode-2 condensate. 
One of the strategies to create a cat state is adiabatically tracking the ground state from the disordered phase (or just above the critical point) to the ordered phase by sweeping the parameter $\Lambda$~\cite{Cirac1998,Yukawa2018}. 
One of the difficulties of this strategy is that it ends in failure due to non-adiabatic transitions unless we take an enough long time~\cite{Caneva2008}, whereas the parity conservation enables us to ignore the small energy gap between the ground state and the first excited state~\cite{Yukawa2018}. 

%
%
\subsection{\label{Sec.STAgen}Superadiabatic scheme}
By using the theory of shortcuts to adiabaticity~\cite{Torrontegui2013}, we can mimic adiabatic dynamics within a short time. 
In the counter-diabatic driving approach~\cite{Demirplak2003,Demirplak2005,Berry2009}, the counter-diabatic Hamiltonian cancels out diabatic changes. 
For a given system described by a time-dependent Hamiltonian $\mathcal{H}_0(t)$, we consider the adiabatic time evolution operator $U_\mathrm{ad}(t)$, i.e., the solution of the Schr\"odinger equation $i\hbar\partial_tU_\mathrm{ad}(t)=\mathcal{H}(t)U_\mathrm{ad}(t)$ becomes adiabatic time evolution of the original Hamiltonian $\mathcal{H}_0(t)$. 
Here, the total Hamiltonian $\mathcal{H}(t)=i\hbar(\partial_tU_\mathrm{ad}(t))U_\mathrm{ad}^\dag(t)$ is decomposed into the original Hamiltonian and the counter-diabatic Hamiltonian
\begin{equation}
\mathcal{H}(t)=\mathcal{H}_0(t)+\mathcal{H}_\mathrm{CD}(t). 
\end{equation}

The approximate counter-diabatic Hamiltonian for the bosonic Josephson junction (\ref{Eq.BJJ}) is given by
\begin{equation}
\mathcal{H}_\mathrm{CD}=\frac{\hbar f}{N}(J_yJ_z+J_zJ_y),
\label{Eq.CDham}
\end{equation}
where the schedule of counter-diabatic driving $f$ is given by
\begin{equation}
f=-\frac{1}{2}\frac{[\hbar\Omega-\frac{\hbar\chi N}{2}(1-\frac{1}{N})]\partial_t(\hbar\Omega)}{[\hbar\Omega-\frac{\hbar\chi N}{2}(1-\frac{1}{N})]^2-(\frac{\hbar\chi N}{2})^2(1-\frac{1}{2N})^2}, 
\end{equation}
in the disordered phase, $0<\Lambda<1$, and
\begin{equation}
f=\frac{[\hbar\chi N(1-\frac{1}{N})(1-\frac{3}{N})+\frac{5\hbar\Omega}{2\Lambda}\frac{1}{N}(1-\frac{7}{4N})]\partial_t(\frac{\hbar\Omega}{2\Lambda})}{[\hbar\chi N(1-\frac{1}{N})-\frac{\hbar\Omega}{2\Lambda}(1-\frac{3}{N})]^2-(\frac{\hbar\Omega}{2\Lambda})^2(1-\frac{1}{2N})^2}, 
\end{equation}
in the ordered phase, $\Lambda>1$, respectively~\cite{Hatomura2018a}. 
Here, for simplicity, we assume that the nonlinear interaction parameter $\chi$ is time-independent and the coupling parameter $\Omega$ is time-dependent. 
With the Hamiltonian
\begin{equation}
\mathcal{H}=\mathcal{H}_\mathrm{BJJ}+\mathcal{H}_\mathrm{CD},
\label{Eq.totham}
\end{equation}
we can create a cat-like state in a bosonic Josephson junction within a relatively short time~\cite{Hatomura2018a}. 
In order to design a continuous schedule of counter-diabatic driving, the time derivative of the coupling parameter should be zero at the initial time, at the final time, and at the time passing the critical point $\Lambda=1$. 

Up to a choice of a reference frame, the counter-diabatic Hamiltonian (\ref{Eq.CDham}) is nothing else but the TACT Hamiltonian as introduced in the context of spin squeezing~\cite{Kitagawa1993} and also studied before as the Lipkin-Meshkov-Glick model \cite{Lipkin1965}.  Note that this form of interaction has not been realized yet, but there are theoretical proposals to implement in experiments~\cite{Opatrny2015,Kajtoch2016}. 
It should be also noted that the similar counter-diabatic Hamiltonian in the thermodynamic limit~\cite{Takahashi2013} and another approximate counter-diabatic Hamiltonian~\cite{Hatomura2017} have also been proposed. 
However, they cannot be applied to the above scheme due to criticality.

%
%
\subsection{\label{Sec.Ploss}Particle losses during generation}

We take particle losses into account by using the Monte Carlo wave function method~\cite{Dalibard1992,Molmer1993}. 
We consider the Schr\"odinger dynamics with the non-Hermitian effective Hamiltonian
\begin{equation}
\mathcal{H}_\mathrm{eff}=\mathcal{H}-\frac{i\hbar}{2}\sum_{n}C_{n}^\dag C_{n},
\label{Eq.effham}
\end{equation}
where $C_{n}$ is the quantum jump operator. 
In this article, we only consider one-body losses for simplicity and thus the quantum jump operator is given by $C_{n}=\sqrt{\gamma_{n}}a_n$, $n=1,2$. 
Here $\gamma_{n}$ gives the decay rate of particles. 
Suppose that a state is given by $|\Psi\rangle$. 
We define the probability of particle losses during a given time interval $\delta t$ as $\delta p\equiv\delta t\sum_{n}\langle\Psi|C_{n}^\dag C_{n}|\Psi\rangle\equiv\sum_{n}\delta p_{n}$. 
With the probability $\delta p$, particles are lost as $|\Psi\rangle\to C_{n}|\Psi\rangle$. 
The type of the particles is selected with the probability $\delta p_{n}/\delta p$. 
Note that we normalize the wave function at each step of numerical simulation. 
We average trajectories in the form of the density operator
\begin{equation}
\rho=\frac{1}{M}\sum_{l=1}^M|\Psi^{(l)}\rangle\langle\Psi^{(l)}|,
\end{equation}
where $|\Psi^{(l)}\rangle$ represents the $l$th trajectory of the stochastic wave function and $M$ represents number of trajectories. 
This density operator $\rho$ for enough large $M$ is equivalent to the solution of the master equation
\begin{equation}
\frac{\partial}{\partial t}\rho=\frac{i}{\hbar}[\rho,\mathcal{H}]+\mathcal{L}\rho,
\label{Eq.Lind}
\end{equation}
where $\mathcal{L}$ is the Lindblad super-operator
\begin{equation}
\mathcal{L}\rho=\sum_{n}\left[C_{n}\rho C_{n}^\dag-\frac{1}{2}(C_{n}^\dag C_{n}\rho+\rho C_{n}^\dag C_{n})\right].
\end{equation}

In this article, we assume the equal decay rate $\gamma\equiv\gamma_{1}=\gamma_{2}$ for simplicity. 
Then, number of lost particles is given by
\begin{equation}
N_\mathrm{loss}=N(1-e^{-\gamma t}),
\end{equation}
where $N$ is number of particles at the initial time, and thus number of residual particles is given by
\begin{equation}
N_\mathrm{res}=N-N_\mathrm{loss}=Ne^{-\gamma t}. 
\label{Eq.exp.decay}
\end{equation}

%
%
\section{\label{Sec.results}Results}
%
%
\subsection{\label{Sec.gen}Generation of cat states}

We can consider two adiabatic schemes. 
One is to start with the trivial ground state in the disordered phase and to track the ground state into the ordered phase~\cite{Cirac1998}. 
Experimentally, we would be able to realize this scheme by cooling down in the disordered phase and by adiabatically sweeping the parameters. 
The other scheme is to start with the ground state just above the critical point~\cite{Yukawa2018}. 
In this case, we do not need to suffer from passing the critical point. 
However, it is difficult to prepare the ground state just above the critical point by cooling down due to the small energy gap. 
We should prepare by using, for example, a $\pi/2$ pulse.

First, we study these two schemes with and without the approximate counter-diabatic Hamiltonian (\ref{Eq.CDham}). 
We assume that the schedule that starts with the ground state in the disordered phase is given by
\begin{equation}
\Lambda^{-1}=48s^5-120s^4+100s^3-30s^2+2,
\label{Eq.schedis}
\end{equation}
where $s=t/t_f$ and $\Lambda^{-1}: 2\to0$. 
This schedule is designed from the boundary conditions, $\Lambda^{-1}(s=0)=2$, $\Lambda^{-1}(s=1/2)=1$, and $\Lambda^{-1}(s=1)=0$ with $\partial_s\Lambda^{-1}|_{s=0,1/2,1}=0$. 
Here, $t_f$ is the time at the end of the generation scheme. 
In contrast, we assume that the schedule that starts with the ground state just above the critical point is given by
\begin{equation}
\Lambda^{-1}=\frac{1}{2}(\cos\pi s+1),
\label{Eq.schecri}
\end{equation}
where $\Lambda^{-1}(s=0)=1$ and $\Lambda^{-1}(s=1)=0$ with $\partial_s\Lambda^{-1}|_{s=0,1}=0$. 
These generation schemes result in the NOON state
\begin{equation}
|\Psi_\mathrm{NOON}\rangle=\frac{1}{\sqrt{2}}(|N,0\rangle+|0,N\rangle), 
\label{Eq.NOONstate}
\end{equation}
if we take enough long time. 
However, due to finite time processes and approximations, our method would result in other states. 
We show how generated states deviate from the NOON state under these situations.

In order to compare them quantitatively, we introduce the residual energy
\begin{equation}
E_\mathrm{res}=E-E_\mathrm{GS}, 
\end{equation}
where $E$ is the energy of the generated state and $E_\mathrm{GS}$ is that of the ground state (the NOON state), and also introduce the quantum Fisher information concerning the operator $J_z$ given by
\begin{equation}
F_Q[\rho,J_z]=2\sum_{\substack{k,l \\ p_k+p_l>0}}\frac{(p_k-p_l)^2}{p_k+p_l}|\langle k|J_z|l\rangle|^2,
\label{eq:fisher-jz}
\end{equation}
where $\rho=\sum_kp_k|k\rangle\langle k|$ is a given density operator. 
For pure states the formula \eqref{eq:fisher-jz} reduces to
\begin{equation}
F_Q[\ket{\Psi},J_z] = 4\langle(\Delta J_z)^2\rangle = 4(\meanv{J_z^2} - \meanv{J_z}^2).
\label{eq:fisher_jz_pure}
\end{equation}

\begin{figure}
\begin{flushleft}
(a)
\end{flushleft}
\vspace{-0.7cm}
\includegraphics[width=8cm]{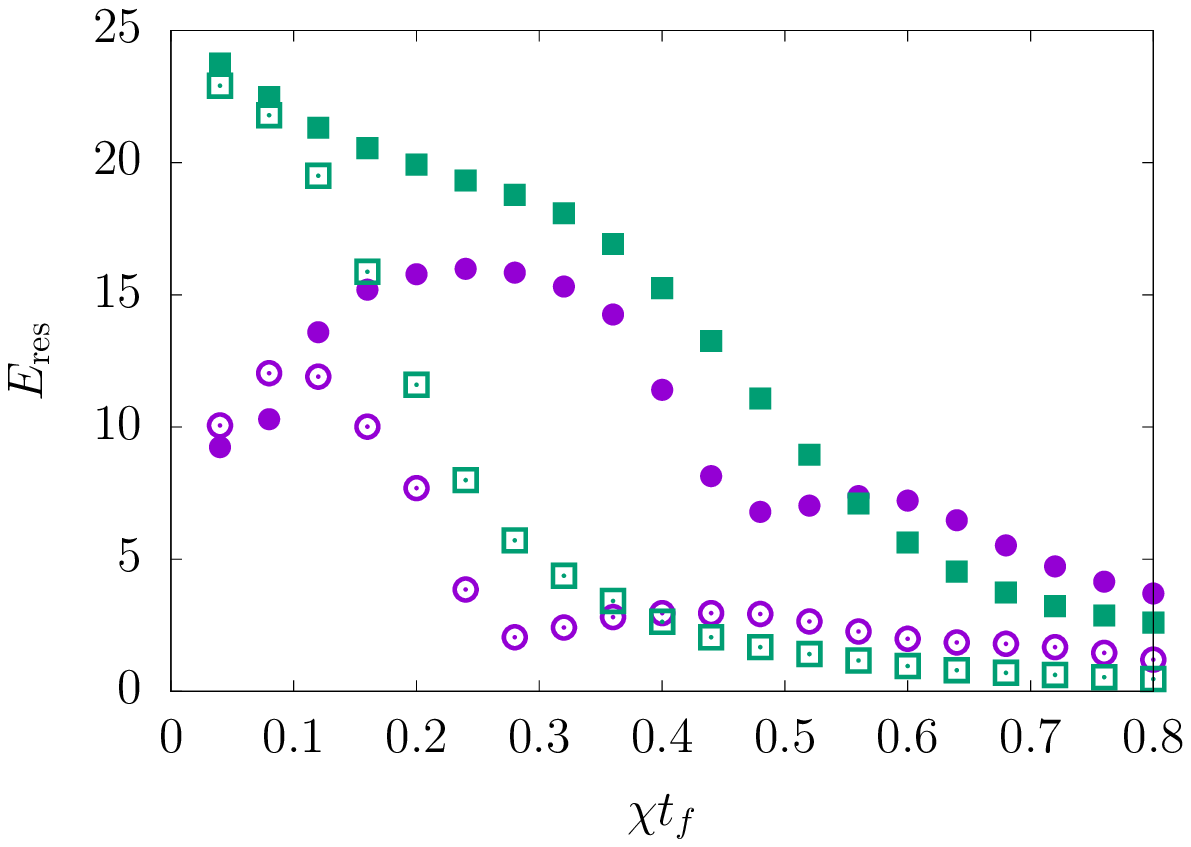}
\begin{flushleft}
(b)
\end{flushleft}
\vspace{-0.7cm}
\includegraphics[width=8cm]{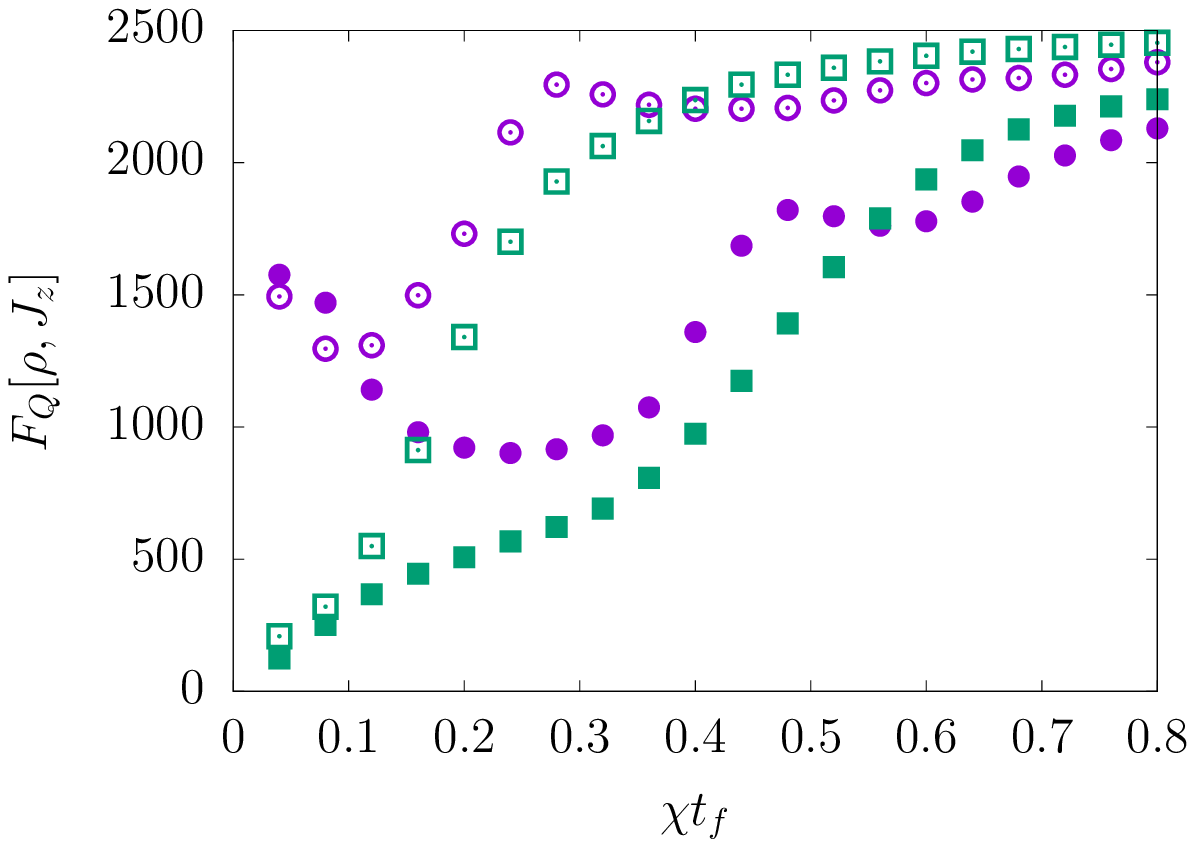}
\caption{\label{Fig.fidqfi} (Color online) (a) Residual energy and (b) quantum Fisher information. In both (a) and (b), the filled symbols represent the schemes starting with the ground state in the disordered phase, the open symbols represent the schemes starting with the ground state just above the critical point, the purple circles represent the schemes assisted by the counter-diabatic Hamiltonian (\ref{Eq.CDham}), and the green squares represent the naive adiabatic schemes. Here, $N=50$. }
\end{figure}

The generated states are close to the ground state when the residual energy is small. 
Moreover, the generated states have large entanglement when the quantum Fisher information is large. 
Especially, the minimum of the residual energy is zero and the maximum of the quantum Fisher information is $N^2$, which can be achieved by the NOON state. 
We plot these quantities of the generated states for both schemes with and without counter-diabatic driving in Fig.~\ref{Fig.fidqfi}. 
Here, $N=50$. 
The filled symbols represent the schemes starting with the ground state in the disordered phase and the open symbols represent the schemes starting with the ground state just above the critical point. 
The purple circles represent the schemes assisted by the counter-diabatic Hamiltonian (\ref{Eq.CDham}) and the green squares represent the naive adiabatic schemes. 
It is clear that adiabaticity is improved by the counter-diabatic Hamiltonian (\ref{Eq.CDham}) in both schemes (\ref{Eq.schedis}) and (\ref{Eq.schecri}) when generation time $\chi t_f$ is small. 
Because we are interested in fast generation schemes, which enable us to minimize effects of decoherence, these results encourage us to investigate short time regimes with the counter-diabatic Hamiltonian.

Hereafter, we only consider the schemes starting with the ground state in the disordered phase because those starting with the ground state just above the critical point are included in the former cases and show similar results.

%
%
\subsection{\label{Sec.pari}Parity conservation and measurement}
In this section, we first discuss about the form of the produced state. 
Because our generation scheme is not the ideal, the generated state is not the NOON state. 
Indeed, we can show that our method generates a superposition of cat states with various sizes
\begin{equation}
|\Psi\rangle=\sum_{m=0}^{N/2}g_m|\Psi_m\rangle, 
\label{Eq.sumsmallcat}
\end{equation}
where, $|\Psi_m\rangle$ is a cat state
\begin{equation}
|\Psi_m\rangle=\frac{1}{\sqrt{2}}(|N-m,m\rangle+|m,N-m\rangle), 
\end{equation}
for $m=0,1,2,\cdots,N/2-1$ and $|\Psi_{N/2}\rangle=|N/2,N/2\rangle$, and thus we call the superposition (\ref{Eq.sumsmallcat}) the ``cats state". 
This is shown by using the parity operator
\begin{equation}
\Pi=\exp[i\pi(J-J_x)], 
\end{equation}
where $J=N/2=(a_1^\dag a_1+a_2^\dag a_2)/2$. 
As the initial state, we prepare the ground state, which has the parity $+1$. 
Then the commutation relations
\begin{equation}
[\mathcal{H}_\mathrm{BJJ},\Pi]=0,\quad[\mathcal{H}_\mathrm{CD},\Pi]=0, 
\end{equation}
ensure conservation of the parity during generation. 
Note that the conservation of the parity during the adiabatic scheme was shown in Ref.~\cite{Yukawa2018}, but it also holds during the superadiabatic scheme. 
Therefore, the parity of the final state is also $+1$, and thus it takes the form of Eq.~(\ref{Eq.sumsmallcat}).

This knowledge of the form of the generated state can be useful to calculate physical quantities. 
Indeed, the residual energy and the quantum Fisher information can be calculated as
\begin{equation}
E_\mathrm{res}=\frac{\hbar\chi}{4}\left[\sum_{m=0}^{N/2}|g_m|^2(N-2m)^2-N^2\right],
\end{equation}
and
\begin{equation}
F_Q[|\Psi\rangle,J_z]=\sum_{m=0}^{N/2}|g_m|^2(N-2m)^2,
\label{Eq.qFisher.cats}
\end{equation}
respectively. 
We can actually reproduce the results in Fig.~\ref{Fig.fidqfi} by using these formulas. 
These expressions also imply that small residual energy leads to large quantum Fisher information, i.e., a low energy state has large entanglement.

Next, we show that the parity operator can also be used to detect the generated cats state, i.e., by measuring the parity after rotation along $z$-axis, we can detect the cats state. 
That is, for a given state $|\Psi\rangle$, we consider rotation $\theta$,
\begin{equation}
|\Psi_\theta\rangle=e^{-i\theta J_z}|\Psi\rangle, 
\end{equation}
and then we measure the parity $\langle\Pi\rangle_\theta=\langle\Psi_\theta|\Pi|\Psi_\theta\rangle$ or its variance $\langle\Delta\Pi^2\rangle_\theta=1-\langle\Pi\rangle_\theta^2$. 
When the given state is the NOON state (\ref{Eq.NOONstate}), these quantities give perfect interference fringes with the frequency $N$~\cite{Bollinger1996}, i.e., $\langle\Pi\rangle_\theta^\mathrm{NOON}=\cos(N\theta)$ and $\langle\Delta\Pi^2\rangle_\theta^\mathrm{NOON}=1-\cos^2(N\theta)=\sin^2(N\theta)$. 
In contrast, when the given state is not a quantum superposed state but a classical mixed state, these quantities are $\langle\Pi\rangle_\theta^\mathrm{cl}=0$ and $\langle\Delta\Pi^2\rangle_\theta^\mathrm{cl}=1$. 
It was shown that a superposition of two coherent spin states also shows similar interference fringes~\cite{Yukawa2018}.

\begin{figure}
\includegraphics[width=8cm]{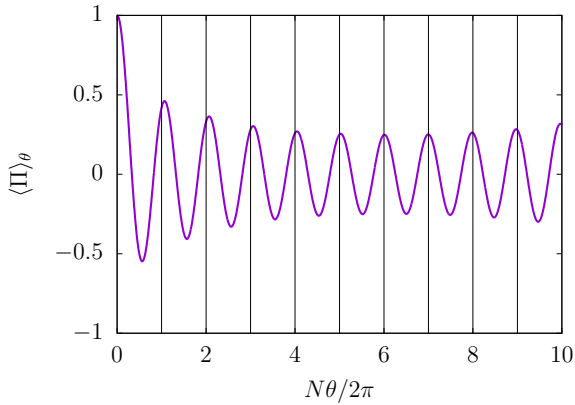}
\caption{\label{Fig.parity} Parity measurement for the cats state generated by the superadiabatic scheme. Here $N=50$ and $|\chi t_f|=0.04$. }
\end{figure}

Our cats state (\ref{Eq.sumsmallcat}) gives
\begin{equation}
\langle\Pi\rangle_\theta=\sum_{m=0}^{N/2}|g_m|^2\cos[(N-2m)\theta], 
\label{Eq.paritysumsmallcat}
\end{equation}
and
\begin{equation}
\langle\Delta\Pi^2\rangle_\theta=1-\left\{\sum_{m=0}^{N/2}|g_m|^2\cos[(N-2m)\theta]\right\}^2, 
\label{Eq.variancesumsmallcat}
\end{equation}
which also show interference fringes. 
However, the frequency of interference fringes is no longer $N$. 
We expect that it is similar to $N$ when $|g_0|^2$ is much larger than others $|g_m|$, $m=1,2,\cdots,N/2$. 
As an example, we depict the result of the parity measurement for a cats state generated by the superadiabatic scheme in Fig.~\ref{Fig.parity}. 
Here, the parameters are $N=50$ and $|\chi t_f|=0.04$. 
We would be able to detect this cats state and to roughly estimate dominant number of particles composing this cats state. 
It would be also possible to determine the distribution $\{|g_m|^2\}$ by using the Fourier analysis.

The measurement of the parity after rotation $e^{-i\theta J_z}$ is nothing but the scheme used in interferometry to estimate $\theta$. 
This interferometric scheme gives one of the lower bounds of quantum Fisher information~\cite{Pezze2009}, which is also known as the error-propagation formula, 
\begin{equation}
F_Q[|\Psi_\theta\rangle,J_z]\ge F_\mathrm{EP}\equiv\frac{|\partial_\theta\langle\Pi\rangle_\theta|^2}{\langle(\Delta\Pi)^2\rangle_\theta}, 
\end{equation}
which can be calculated by using Eqs.~(\ref{Eq.paritysumsmallcat}) and (\ref{Eq.variancesumsmallcat}). 
In the limit $\theta\to0$, it leads to
\begin{equation}
F_\mathrm{EP}\to\sum_{m=0}^{N/2}|g_m|^2(N-2m)^2, 
\end{equation}
which is identical to the quantum Fisher information~(\ref{Eq.qFisher.cats}). 
That is, the parity measurement with rotation $e^{-i\theta J_z}$ is the best way of interferometry that maximally extracts the potential of the cats state (\ref{Eq.sumsmallcat}). 
This agrees with the known properties of the path symmetric states in optics~\cite{Seshadreesan2013}.

%
%
\subsection{\label{Sec.dist}Distributions of states and particle losses}
Now we study effects of particle losses during adiabatic and superadiabatic generation of the cats state. 
We average 5000 trajectories of the Monte Carlo wave function, which is enough large to describe loss processes. 

Here we utilize the SU(2) Wigner function~\cite{Dowling1994} in order to visualize the distributions of the produced states and their entanglement. 
The SU(2) Wigner function is usually defined for a fixed number of particles. 
However, in our case, the density operator is block-diagonalized for each number of particles as
\begin{equation}
\rho = \sum_{n=0}^{N}p_n\rho_{n},
\end{equation}
due to particle losses. 
Here, $p_n$ is a probability of finding $n$ atoms and $\rho_n$ is the density operator in the $n$ particles subspace. 
In this case, the SU(2) Wigner function takes the following form
\begin{equation}
W(\theta,\phi) = \sum_{n=0}^{N}p_nW_n(\theta,\phi),
\end{equation}
where $W_n(\theta,\phi)$ is the SU(2) Wigner function for the density operator in the $n$ particles subspace~\cite{Dowling1994}.

\begin{figure*}
\includegraphics[width=7cm]{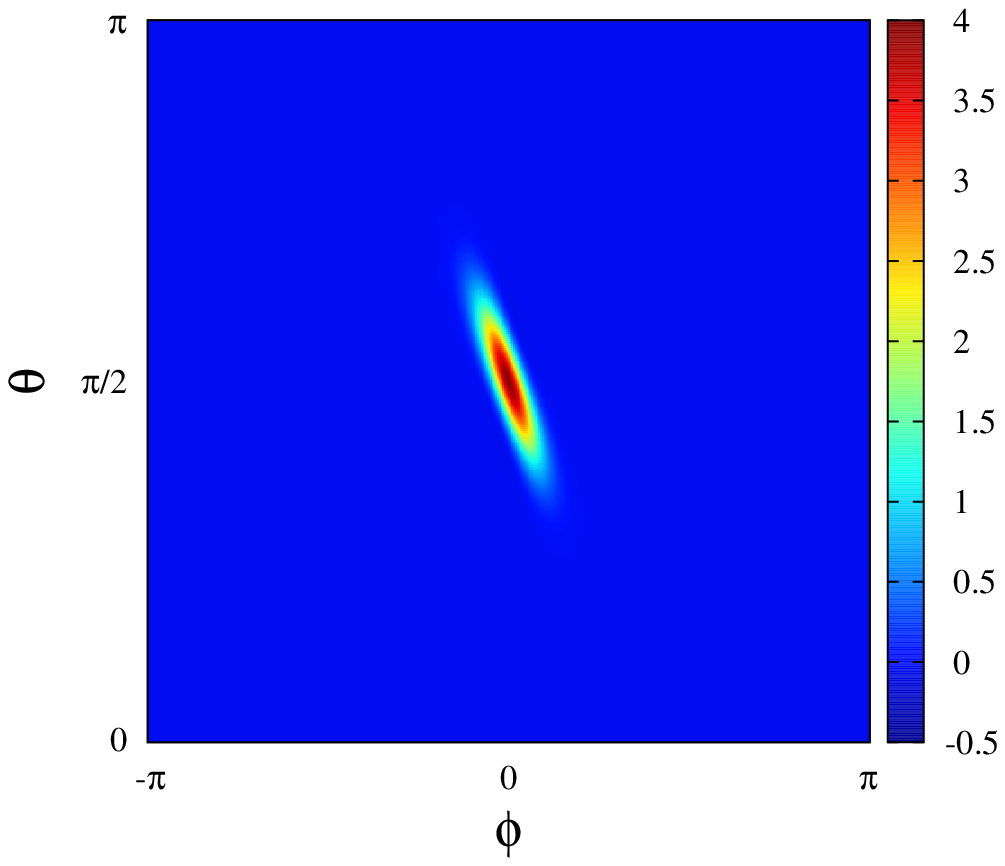}
\includegraphics[width=7cm]{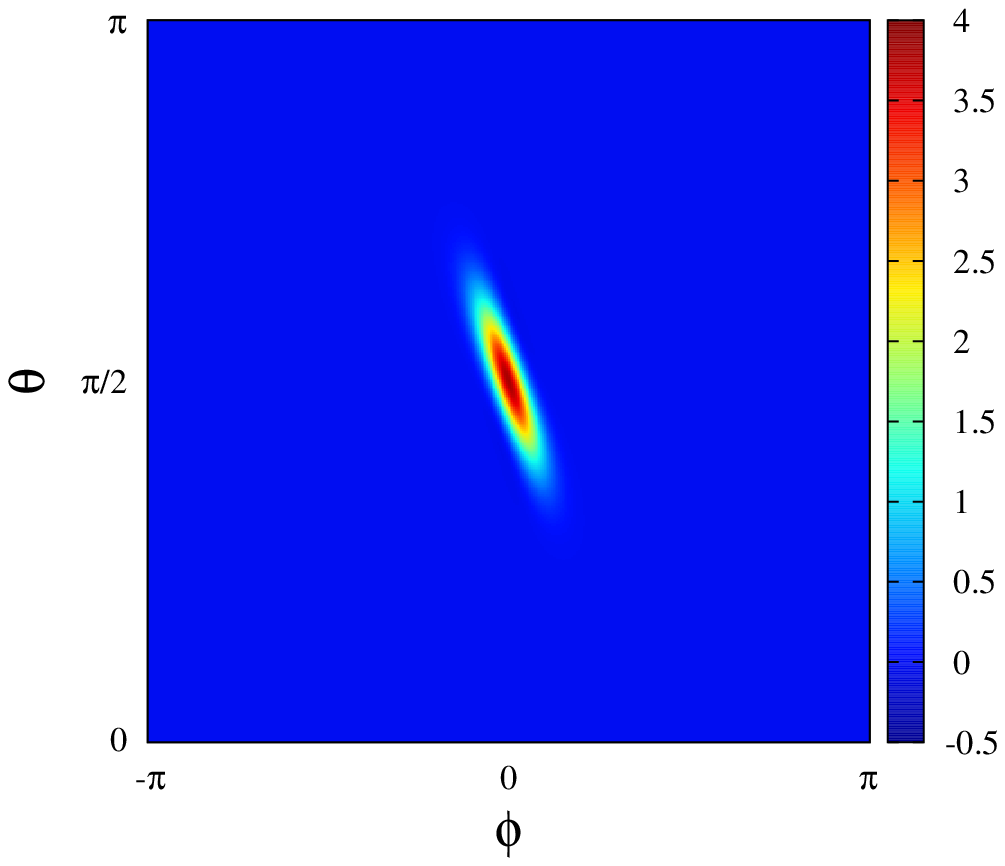}
\includegraphics[width=7cm]{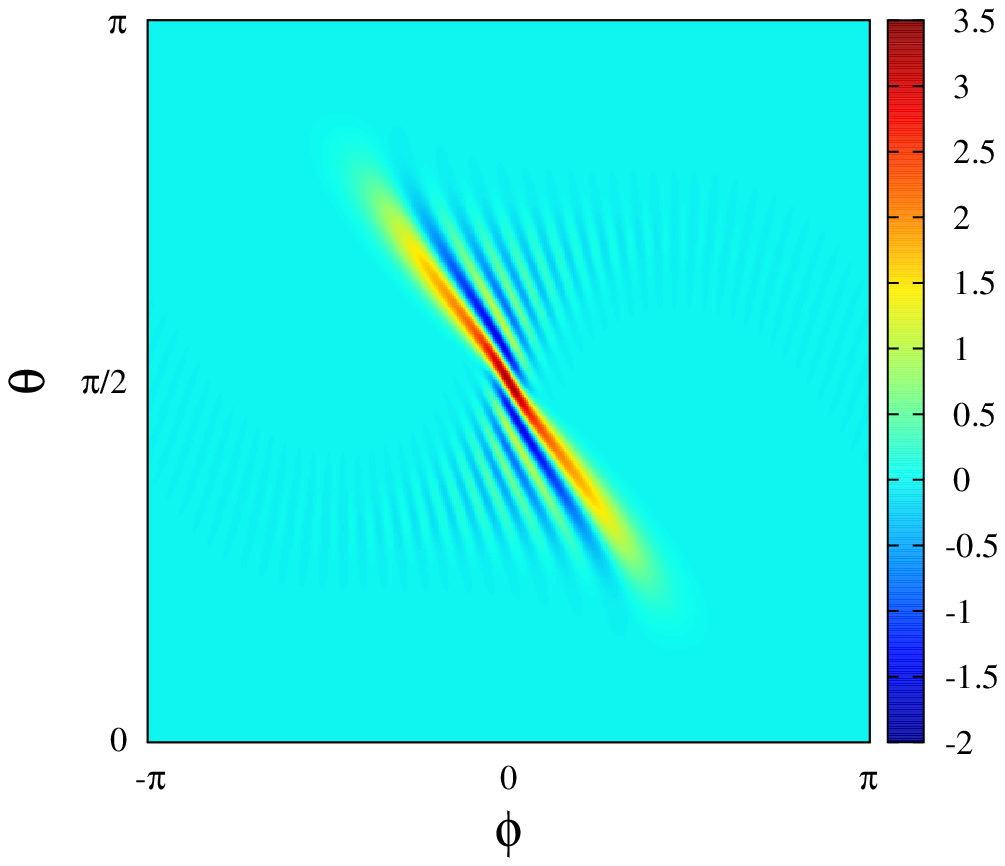}
\includegraphics[width=7cm]{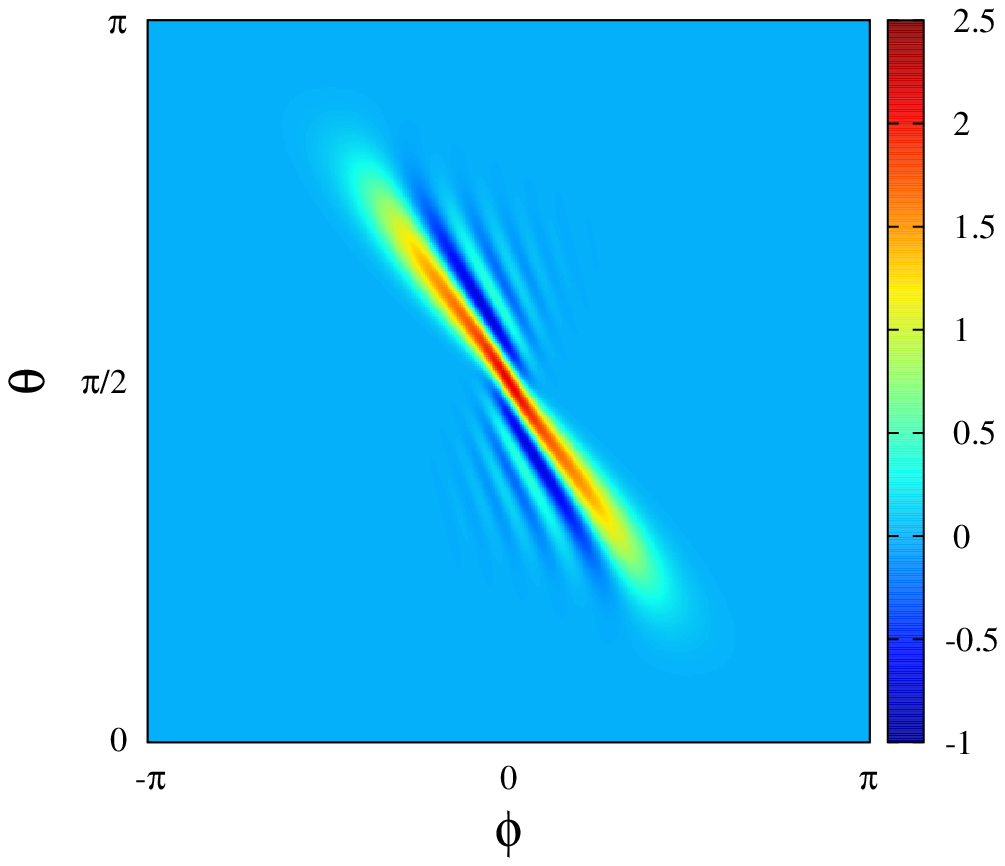}
\includegraphics[width=7cm]{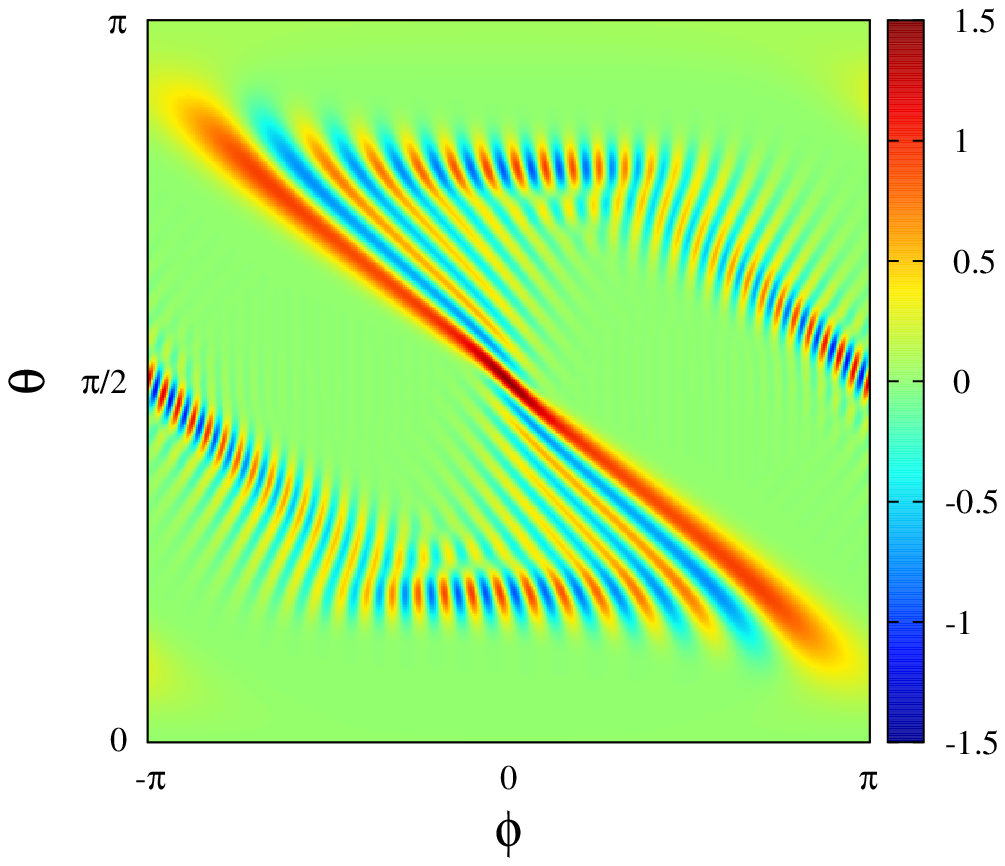}
\includegraphics[width=7cm]{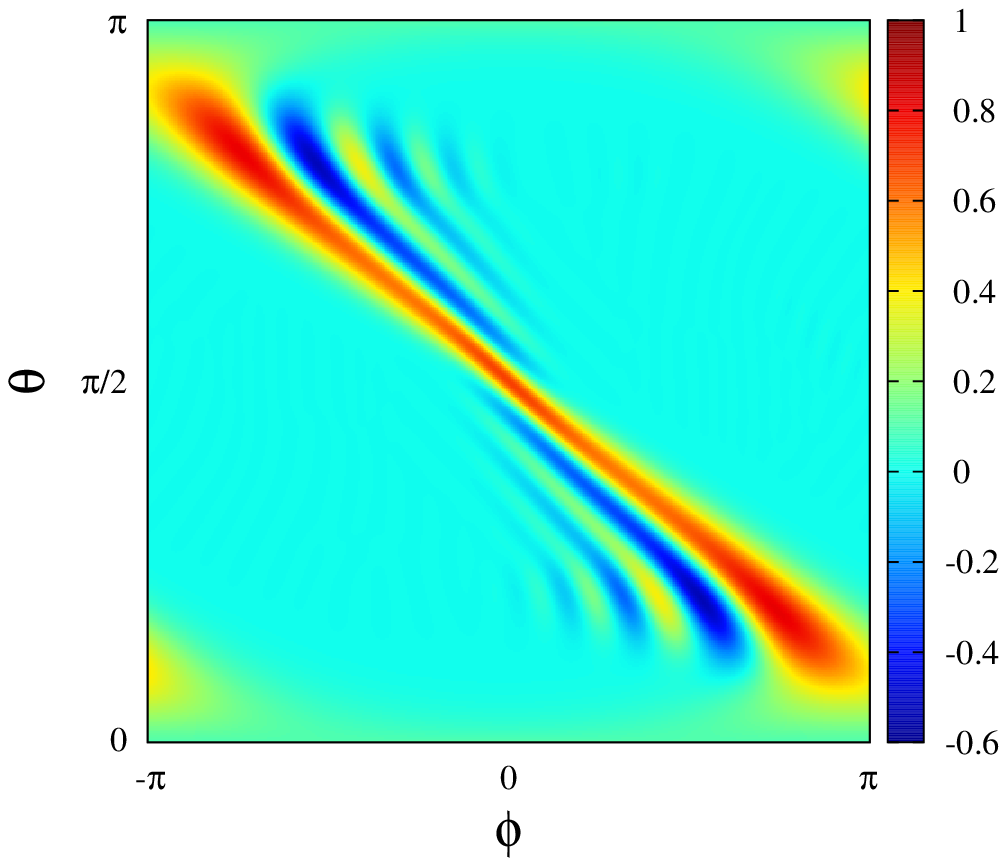}
\caption{\label{Fig.Wigner-ncd} (Color online) The Wigner functions of states generated by the adiabatic scheme. The left panels are the cases without particle losses and the right panels are those with particle losses ($\gamma=0.05$). The upper panels are generated within $\chi t_f=0.04$, the middle panels are within $\chi t_f=0.2$, and the lower panels are $\chi t_f=0.4$. Here, $N=50$. }
\end{figure*}

In Fig.~\ref{Fig.Wigner-ncd}, we depict the SU(2) Wigner function of states generated by the naive adiabatic scheme. 
The left panels are the cases without particle losses and the right panels are those with particle losses ($\gamma=0.05$). 
The upper panels are generated within $\chi t_f=0.04$, the middle panels are within $\chi t_f=0.2$, and the lower panels are $\chi t_f=0.4$. 
Note that numbers of lost particles are about five percent of the initial particles for $\chi t_f=0.04$, about twenty two percent of those for $\chi t_f=0.2$, and about thirty nine percent of those for $\chi t_f=0.4$, respectively. 
When the generation time is too short ($\chi t_f=0.04$), the state cannot follow change of the Hamiltonian, and thus the final state results in a spin coherent-like state, which is similar to the initial state. 
For much longer duration of evolution ($\chi t_f=0.2$), the final state becomes a spin squeezed-like state. 
When we take a long time ($\chi t_f=0.4$), the final state is much more squeezed. 
We have to take much more time in order to obtain a cat-like state.

\begin{figure*}
\includegraphics[width=7cm]{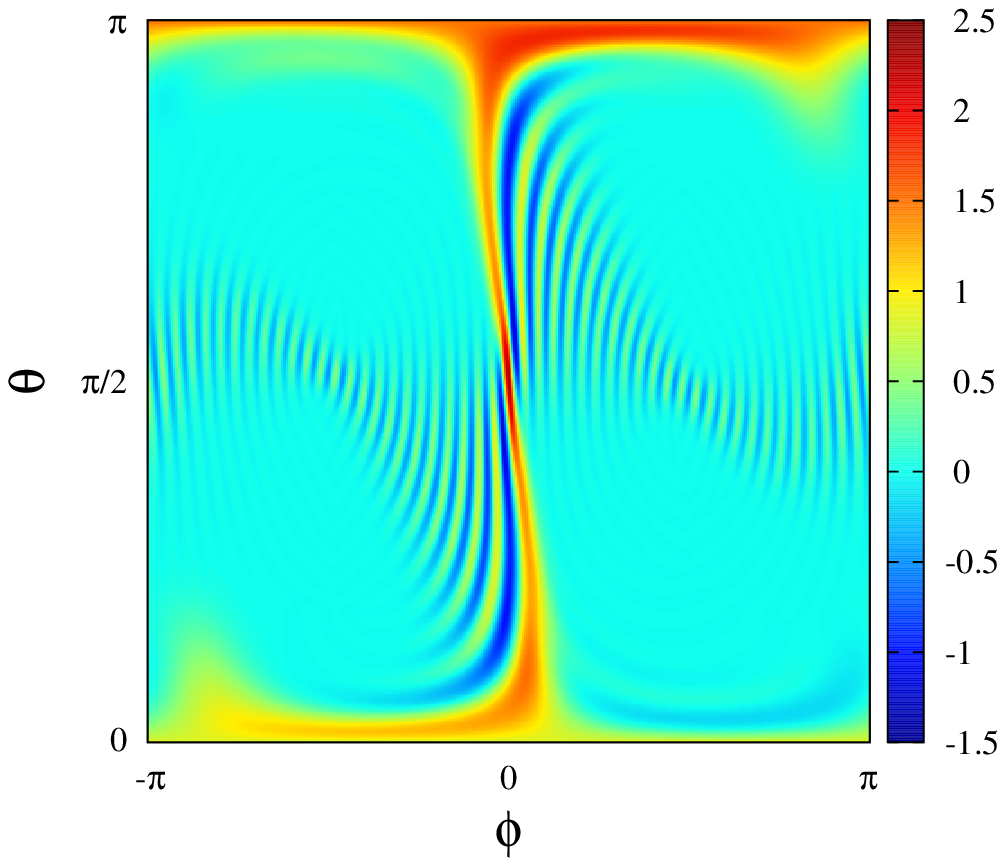}
\includegraphics[width=7cm]{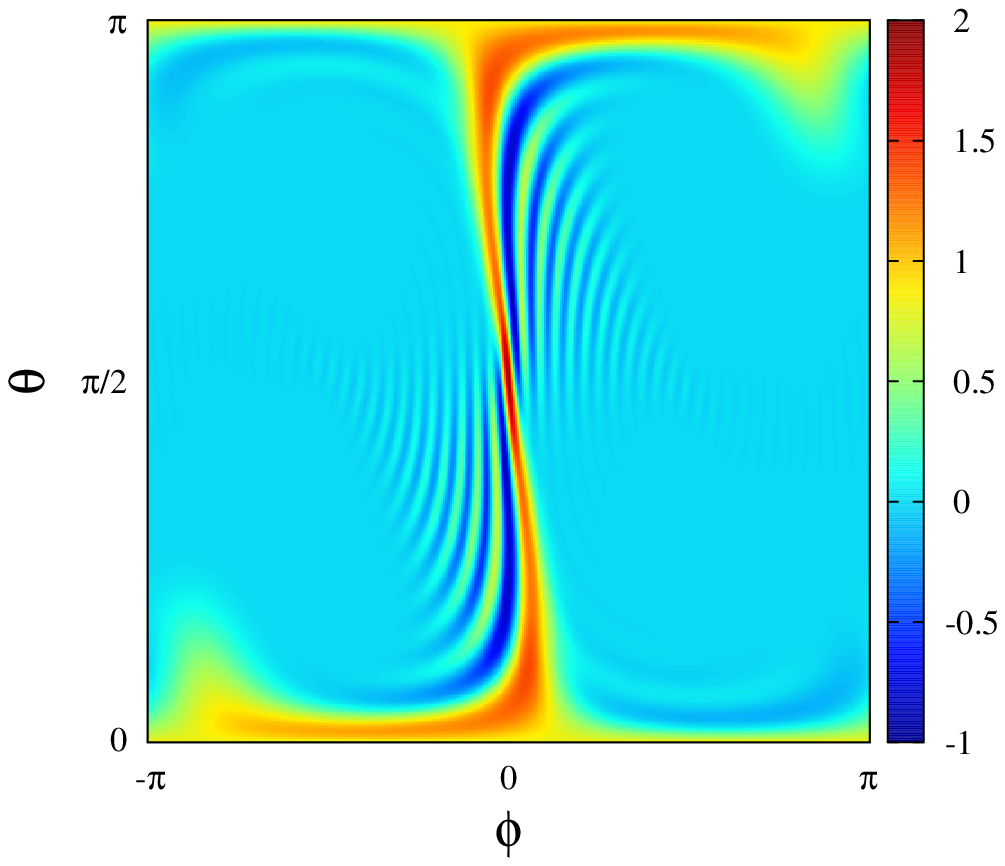}
\includegraphics[width=7cm]{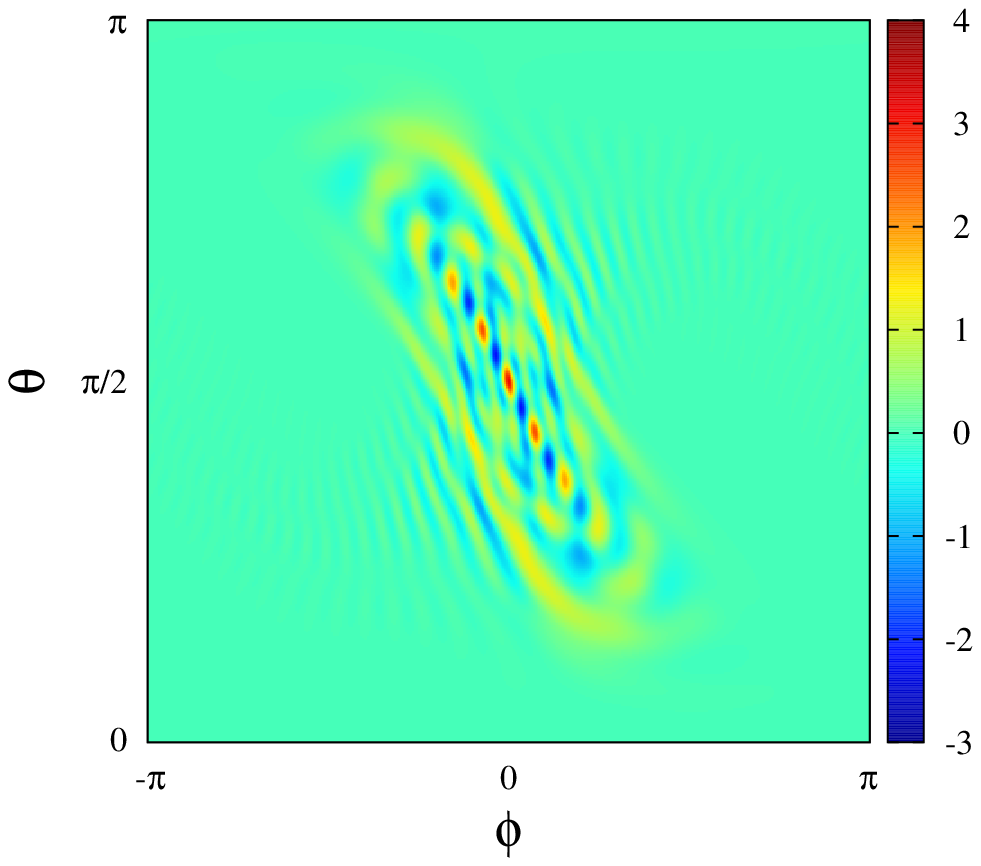}
\includegraphics[width=7cm]{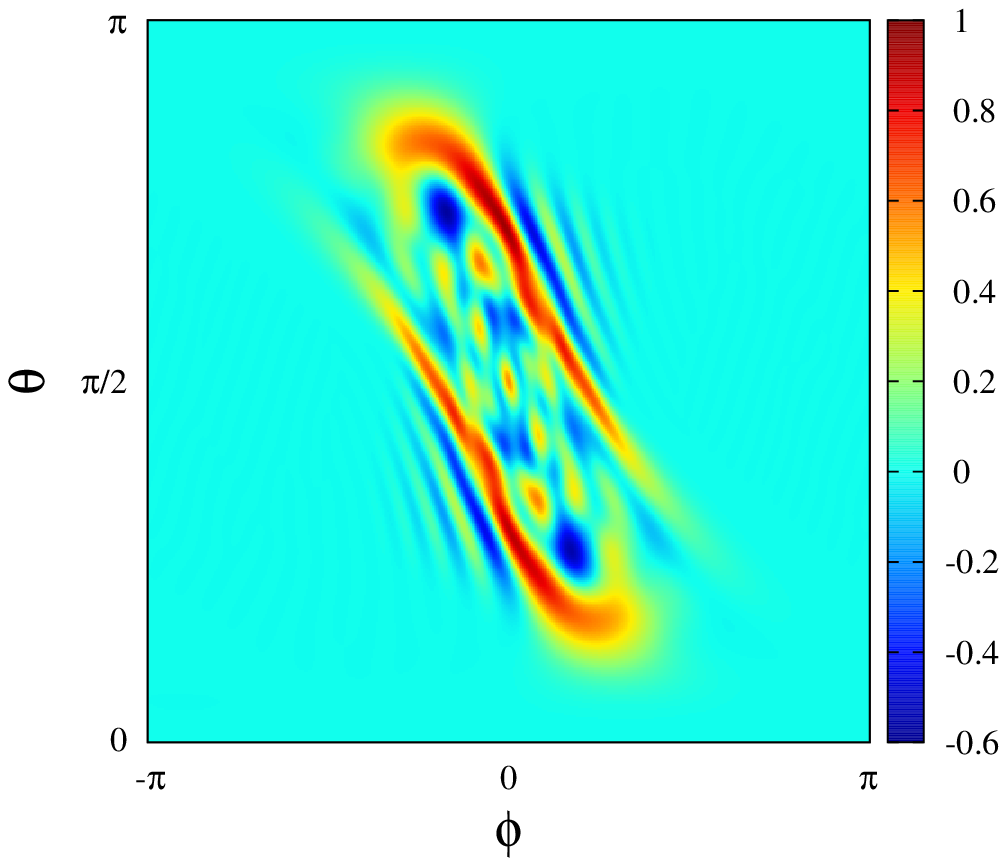}
\includegraphics[width=7cm]{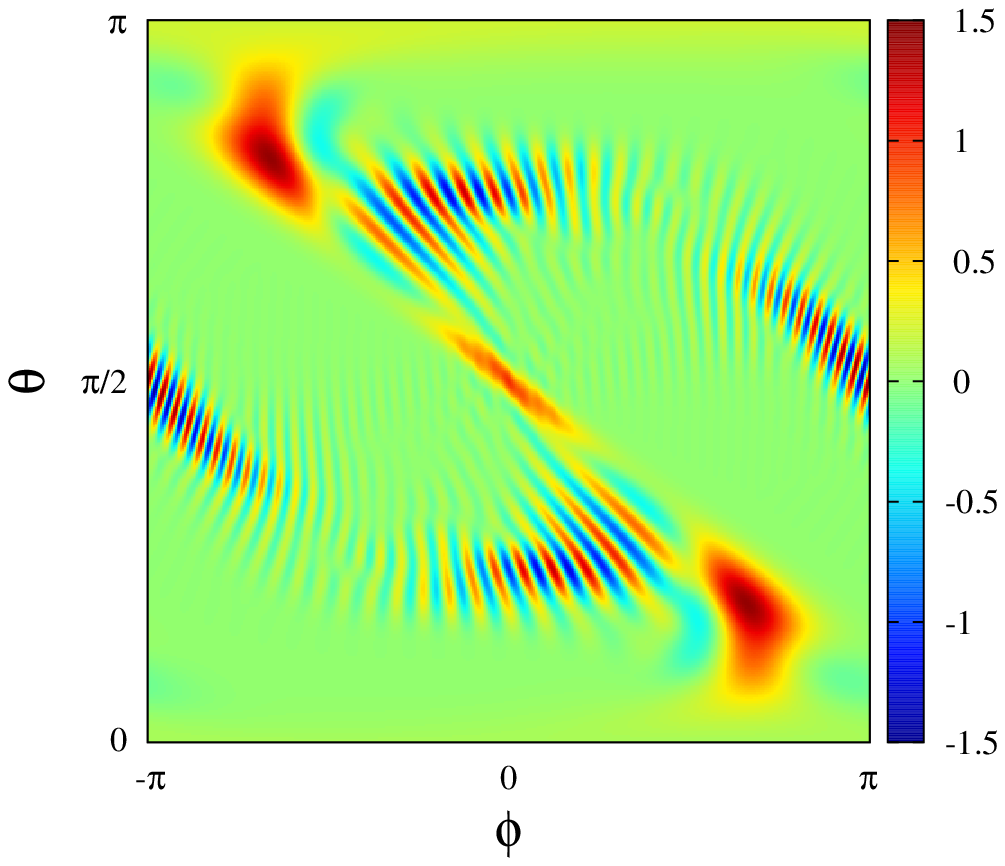}
\includegraphics[width=7cm]{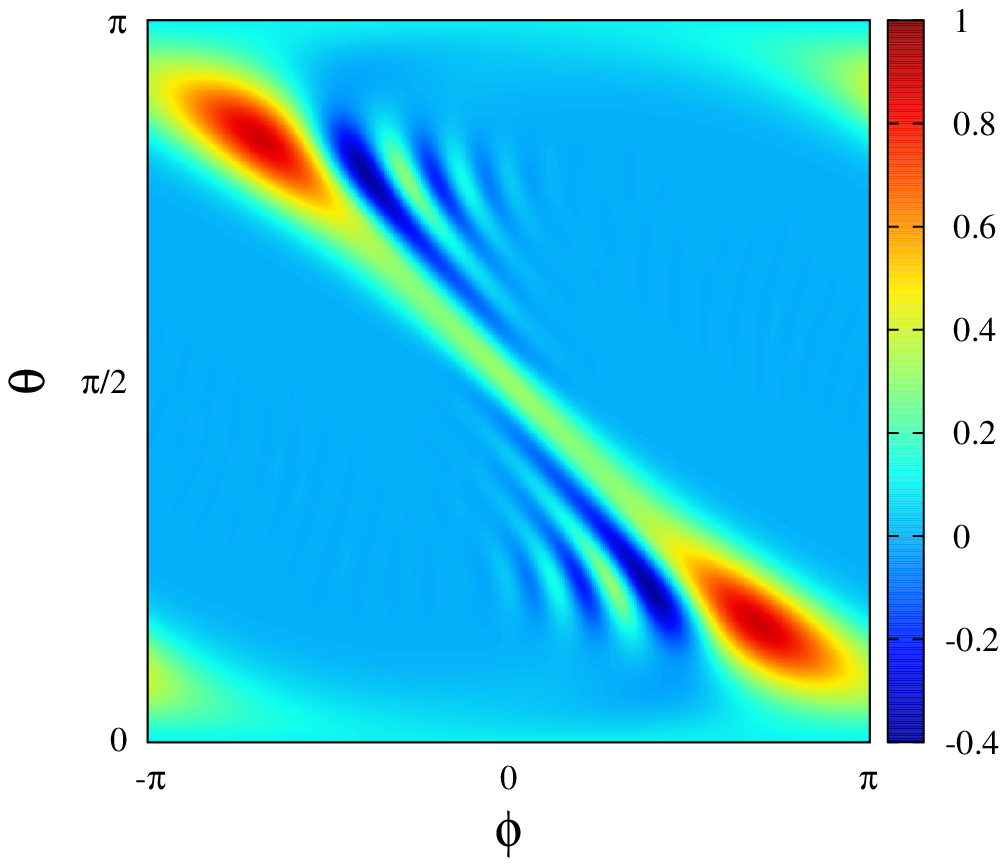}
\caption{\label{Fig.Wigner} (Color online) SU(2)Wigner functions of states generated by the superadiabatic scheme. The left panels are the cases without particle losses and the right panels are those with particle losses ($\gamma=0.05$). The upper panels are generated within $\chi t_f=0.04$, the middle panels are within $\chi t_f=0.2$, and the lower panels are $\chi t_f=0.4$. Here, $N=50$. }
\end{figure*}

In Fig.~\ref{Fig.Wigner}, we depict the SU(2) Wigner function of states generated by the superadiabatic scheme. 
The parameters are similar to Fig.~\ref{Fig.Wigner-ncd}. 
Even if the generation time is short ($\chi t_f=0.04$), two peaks appear with large distance, but with a ``tail" between them. 
Therefore, it rather resembles an intermediate state between a cat state and a spin squeezed state. 
Then at the generation time $\chi t_f=0.2$, it is clearly a complicated non-Gaussian state, whereas for $\chi t_f =0.4$ one obtains the SU(2) Wigner function with two distinct local maxima, which is similar to that of a cat state.

In the superadiabatic scheme, as well as in the adiabatic scheme, particle losses do not change distributions drastically despite the fact that until  $\chi t_f=0.4$  around 20 particles from initial 50 particles have been lost. 
Apparently the state is changed by losses, but still there are thin negative fringes in the SU(2) Wigner function, which indicate entanglement between atoms. 
Moreover, thin negative fringes in the SU(2) Wigner function also suggest potential usefulness for high interferometric sensitivity. 
This stems from the fact that an interferometer is a device which rotates a state around the interferometric axis. 
This device could be more sensitive if smaller rotation angles can be distinguished with this state. 
Therefore, thinner structures in the SU(2) Wigner function lead to more chance for precise interferometry. 
Of course, more detailed analysis is necessary to confirm the interferometric usefulness. 
In order to that, we will come back to the parity measurement and also behavior of the quantum Fisher Information under particle losses.

%
%

\subsection{\label{Sec.surv}Survival of cats states under particle losses}
In this section, we see how entanglement properties and outcomes of the parity measurement change when particle losses take place.

As discussed in Sec.~\ref{Sec.pari}, the parity conservation is the key concept of adiabatic and superadiabatic generation. 
However, the parity is no longer conserved quantity when particle losses take place. 
Indeed, the Lindblad equation (\ref{Eq.Lind}) leads to
\begin{equation}
\frac{\partial}{\partial t}\langle\Pi\rangle=2\gamma\langle \bb{J_x-J}\Pi\rangle, 
\label{eq:dynamics-parity}
\end{equation}
which shows time evolution of the parity. 
Because $(J_x-J)$ and $\Pi$ commute, we obtain
\begin{equation}
\mathrm{Tr}\{(J_x-J)[\Pi,\rho]\}=0. 
\end{equation}
This leads to the invariance of the density operator under the parity operation
\begin{equation}
\rho=\Pi\rho\Pi. 
\end{equation}
It ensures that the density operator is block-diagonalized in the parity $\Pi=+1$ sector and the parity $\Pi=-1$ sector, i.e., the density operator can be written as
\begin{equation}
\rho=\rho^++\rho^-. 
\end{equation}
Here we introduce the parity $\Pi=\pm1$ eigenstates
\begin{equation}
|\Psi_{m,\pm}^{(M)}\rangle=\frac{1}{\sqrt{2}}(|M-m,m\rangle\pm|m,M-m\rangle), 
\end{equation}
and then $\rho^\pm$ can be written as
\begin{equation}
\rho^\pm=\sum_{M=0}^N\sum_{m,n=0}^{M/2}g_m^\pm(g_n^\pm)^\ast|\Psi_{m,\pm}^{(M)}\rangle\langle\Psi_{n,\pm}^{(M)}|. 
\end{equation}

\begin{figure}
\includegraphics[width=8cm]{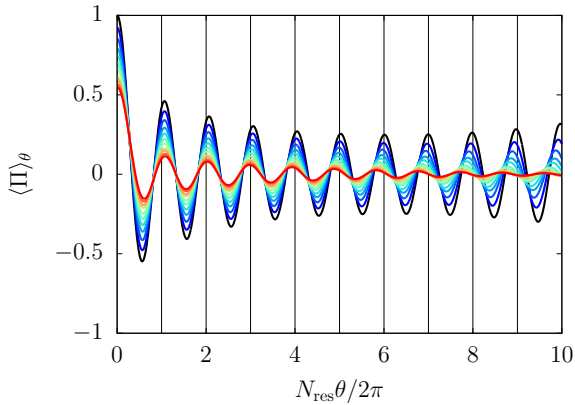}
\caption{\label{Fig.pariloss} (Color online) Parity measurement for states generated by the superadiabatic scheme under particle losses with $\gamma=0, 0.01, 0.02, \cdots, 0.1$ (plotted by gradation from black to red). Here $N=50$ and $|\chi t_f|=0.04$. }
\end{figure}

Then, the parity after rotation $e^{-i\theta J_z}$ is given by
\begin{equation}
\langle\Pi\rangle_\theta=\sum_{M=0}^N\sum_{m=0}^{M/2}(|g_m^+|^2-|g_m^-|^2)\cos[(M-2m)\theta]. 
\end{equation}
Therefore, we still have a chance to observe interference fringes. 
Indeed, we can find them in Fig.~\ref{Fig.pariloss}, where the outcomes of the parity measurement are plotted for $\gamma=0, 0.01, 0.02, \cdots, 0.1$ (percentage of numbers of lost particles with respect to those of initial particles is from zero percent to about ten percent). 
When large number of particles are lost, we cannot easily find number of residual particles composing the cats state, which is associated with blurred distributions. 
However, it should be enough to find the evidence of the cats state.

\begin{figure}
\includegraphics[width=8cm]{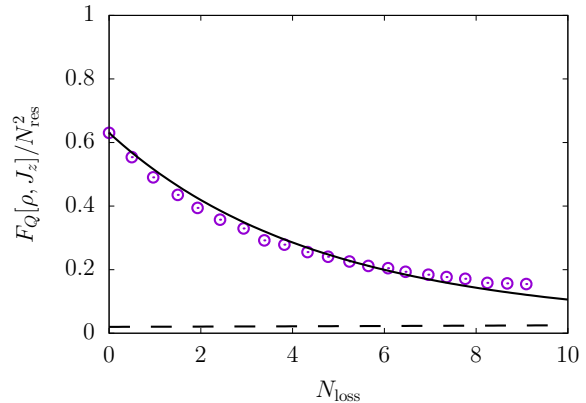}
\caption{\label{Fig.qfiloss} Quantum Fisher information scaled by the square of number of residual particles $N_\mathrm{res}$ with respect to the number of lost particles $N_\mathrm{loss}$. Number of the initial particles is given by $N=50$. Here, $|\chi t_f|=0.04$. The Heisenberg limit is given by $F_Q[\rho,J_z]/N_\mathrm{res}^2=1$ and the standard quantum limit is given by $F_Q[\rho,J_z]/N_\mathrm{res}^2=1/N_\mathrm{res}=1/(N-N_\mathrm{loss})$, which is represented by the dashed curve. A possible fitting $F_Q(N_\mathrm{loss})/N_\mathrm{res}^2\sim [F_Q(0)/N_\mathrm{res}^2]\exp[-\gamma N_\mathrm{res}t_f/4]=[F_Q(0)/(N-N_\mathrm{loss})^2][(N-N_\mathrm{loss})/N]^{(N-N_\mathrm{loss})/4}$ is also given by the solid curve. }
\end{figure}

Next, we study how the quantum Fisher information decreases when some particles are lost. 
Because the quantum Fisher information of cat states is scaled by the square of number of particles, we plot $F_Q[\rho,J_z]/N_\mathrm{res}^2$ with respect to $N_\mathrm{loss}$ in Fig.~\ref{Fig.qfiloss}. 
The quantum Fisher information remains above the standard quantum limit even if about twenty percent of the initial particles are lost. 
It implies survival of entanglement. 
A possible fitting of the quantum Fisher information with respect to a number of lost particles is $F_Q(N_\mathrm{loss})\sim F_Q(0)\exp[-\gamma N_\mathrm{res}t_f/4]=F_Q(0)[(N-N_\mathrm{loss})/N]^{(N-N_\mathrm{loss})/4}$, which is also plotted in Fig.~\ref{Fig.qfiloss}. 
Here we use Eq.~(\ref{Eq.exp.decay}) in the second equality. 
This is quite unexpected scaling. Indeed, in the case of the paradigmatic one-axis twisting Hamiltonian with symmetric losses, the quantum Fisher Information of the target cat state decreases as  $F_Q(0)\exp[-\gamma N t_f]$. 
The present case also shows the exponential decay but it depends on a number of residual atoms (and also has a factor $1/4$) instead of a number of the initial atoms, and thus it gives slower decay.

%
%
\section{\label{Sec.sum}Summary}
We showed that the superadiabatic scheme can generate low energy and highly entangled states within a short time compared with the naive adiabatic scheme. 
The resulting state is not the NOON state but the ``cats state", which is a superposition of the cat states with different sizes, due to the approximation in counter-diabatic terms. 
The form of this cats state (\ref{Eq.sumsmallcat}) is ensured by the parity conservation of the bosonic Josephson junction Hamiltonian and of the approximate counter-diabatic Hamiltonian. 
This cats state could be detected by the parity measurement after rotation along $z$-axis. 
Dominant size of the cat states in the cats state can be estimated from the frequency of the parity measurement if the distribution of the size of the cat states is sharp enough. 
The Fisher information estimated from the error-propagation formula for the parity measurement coincides with the quantum Fisher information, and thus the parity measurement is not only able to detect the cats state but also it is the best way of interferometry that maximally extracts potential of the cats state. 

We also investigated influence of particle losses during superadiabatic generation of the cats state. 
Particle losses blur the distributions of states, but they do not drastically change those. 
In particular, negative interferometric fringes in the SU(2) Wigner function survive even if some particles are lost, which is evidence of entanglement and of potential usefulness. 
We also showed a possibility to detect the cats state by using the parity measurement even if some particles are lost. 
Although the quantum Fisher information decreases as the result of particle losses, i.e., the size of entanglement becomes small, it still remains above the standard quantum limit even if a large amount of particles is lost. 
Note that the parity measurement is no longer the best way of interferometry if particle losses take place. 
It is a future work to study how to extract remaining usefulness of the cats state.

Here we studied with small number of particles $N=50$ due to a problem of the computational cost. 
One would wonder if the discussion in the present article works for larger number of particles or not. 
We observed that the quantum Fisher Information concerning $J_z$ scales as $F_Q(N_\mathrm{loss})\sim F_Q(0)e^{- \gamma N_\mathrm{res}t_f/4}$.
Given the typical rate of one-body losses around $\gamma = 0.1$~Hz (see for instance \cite{Stamper-Kurn1998, Gross2010, Whitlock2010, Ockeloen2013}) and typical timescales in the ultracold atoms experiments around $10$~ms, we expect that the number of atoms, for which the scheme could beat the standard quantum limit in interferometry, would be several hundreds or a few thousands of atoms. 
Indeed, it is known that the time required to implement the naive adiabatic scheme is comparable to the coherence time~\cite{Yukawa2018} when we consider, for example, hundreds of trapped ions~\cite{Bohnet2016}. 
In order to create macroscopically entangled states with high quality and with high probability, and also to utilize those after creation, generation scheme should be much faster than coherence time. 
We expect that the interaction in the form of Eq.~(\ref{Eq.CDham}) will be realized in the near future, for example, by implementing the theoretical proposal~\cite{Opatrny2015,Kajtoch2016}, and that our scheme will be a strong candidate to create macroscopically entangled states in bosonic Josephson junctions.


\begin{acknowledgments}
TH is supported by the Ministry of Education, Culture, Sports, Science and Technology of Japan through the Elements Strategy Initiative Center for Magnetic Materials and supported by the Japan Society for the Promotion of Science (JSPS) through the Research Fellowship for Young Scientists (DC2) and through the Program for Leading Graduate Schools: Material Education program for the future leaders in Research, Industry, and Technology of the University of Tokyo. 
This work is supported by JSPS KAKENHI Grant Number JP18J11053. 
KP is supported by the (Polish) National Science Center Grant 2014/13/D/ST2/01883.
\end{acknowledgments}

\bibliography{lossbib,lossbib2}

\end{document}